# Framework for lung CT image segmentation based on UNet++

**Hao Ziang[1,3], Jingsi Zhang[2,4], Lixian Li[3,5]**

[1] Cardiff University, UK
[2] Khoury College of Computer Sciences, Northeastern University, Boston, MA, USA
[3] Florida College of Integrative Medicine, Orlando, Florida, USA

[3] LeonHao214@outlook.com
[4] zhang.jings@northeastern.edu
[5] lli1@fcim.edu

**Abstract.** Recently, the state-of-art models for medical image segmentation is U-Net and their variants. These networks, though succeeding in deriving notable results, ignore the practical problem hanging over the medical segmentation field: overfitting and small dataset. The over-complicated deep neural networks unnecessarily extract meaningless information, and a majority of them are not suitable for lung slice CT image segmentation task. To overcome the two limitations, we proposed a new whole-process network merging advanced UNet++ model. The network comprises three main modules: data augmentation, optimized neural network, parameter fine-tuning. By incorporating diverse methods, the training results demonstrate a significant advantage over similar works, achieving leading accuracy of 98.03% with the lowest overfitting. potential. Our network is remarkable as one of the first to target on lung slice CT images.

## 1. Introduction

Statistics presented on Global Cancer Statistics highlight the widespread prevalence of lung cancer as a significant health concern. Contemporarily, tumor resection surgery remains the mainstream of lung cancer treatment, and X-ray, as one of the most common image-assisted means, plays a vital role for clinicians to diagnose and assess conditions. However, tradition manual segmentation lacks efficacy and comes along inevitable errors due to complexity within lungs and huge workload of physicians. Radiation oncologists showed a very skewed distribution globally due to high demand for equipment, physician, and radiation planning expertise [1]. Therefore, it is necessary to manipulate computer science for the segmentation of specific areas and extraction of features in medical images.

    Recent researches in the fields of computer vision allow for computer-aided diagnosis and smart medicine. Convolutional neural networks (CNNs) have proven effective in accurately segmenting CT images. This success is credited to the capability of CNNs in learning a hierarchical representation of raw input data, and its introduction significantly propelled the advancement of various segmentation models. In 2015, Long et al. [2] pioneered fully convolutional networks (FCNs), followed by the subsequent introduction of UNet by Ronneberger et al. [3]. Both frameworks share a fundamental key concept: skip connections. In FCN, up-sampled feature maps are combined with feature maps skipped

from the encoder, whereas UNet concatenates them and adds convolutions and non-linearities between each up-sampling step. This encoder-decoder architecture merges deep, semantic, coarse-grained feature maps from the decoder sub-network with shallow, low-level, fine-grained feature maps from the encoder sub-network. This approach has demonstrated effectiveness in recovering fine-grained details of target objects [4] even against complex backgrounds [5]. Arguably, image segmentation in natural images has gone beyond satisfactory, but meeting the much stricter requirements on the medical images remains a primary challenge.

Medical images require precise segmentation due to critical but tiny details, where even minor errors can lead to severe medical consequences. CT images, with limited color information and 2D complexity, often face overfitting challenges during training, reducing the efficacy of traditional models in practical scenarios [6].

To address these issues, this paper introduces a UNet++-based framework for targeted image segmentation. Our contributions include: (1) optimizing the overall architecture, incorporating advanced pre-processing and parameter control techniques to mitigate overfitting; (2) integrating recent model improvements into UNet++ for enhanced CT image segmentation; and (3) employing data augmentation to address small dataset limitations, improving UNet++ generalization for pneumonia diagnosis. This approach ensures higher precision and reliability in medical image segmentation.

## 2. Related work

In medical image segmentation, the first powerful encoder-decoder structure, UNet proposed by Ronneberger et al. [3], has soon become the mainstream. Its advanced encoder-decoder architecture, perfect symmetric structure, and skip connection solve the general problems of CNN networks. However, the utilization of skip connection design in an encoder-decoder network suffers from the large semantical distinction. To recede the fusion of semantically dissimilar feature from plain skip connections in UNet, several enhancements have been proposed by researchers. Attention UNet [7] directs attention to the salient features of the encoder and decoder counterpart through attention gate. MultiResUNet [8] mitigates the semantic discrepancy between encoder and decoder by incorporating Res-Path as a skip connection between them. UNet++ [9] further reinforces these connections by introducing nested and dense skip connections, allowing the network to more flexibly exploit the features from various levels and thus narrowing the semantic gap between the encoder and decoder. The adoption of UNet++ has led to significant advancements in a variety of segmentation tasks.

However, the above networks all suffer a potential risk of overfitting. Summarizing the previous experience, M.Z. Khan suggested random weight initialization, dropout function, ensemble methods, and transfer learning as strategies to mitigate the issue of overfitting [10]. Non-network structural factors may also play a critical role in improving segmentation results. Isensee et al. [11] argued that excessive manual tweaking of network architecture could potentially lead to overfitting on a given dataset. They, therefore, proposed a medical image segmentation framework no-new-UNet (nnUNet), focusing on the stage of pre-processing stages, training procedures, inference methods, and post-processing techniques. Whereas in practical application, the experience-based improvement of network designs lack of adequate interpretability supporting theory. Overly complex network architecture also suggests a higher overfitting risk. With concern to the issue, Weng et al. [12] proposed an NAS-UNet, containing two same architecture DownSC and UpSC, for medical image segmentation tasks. The NAS-UNet outperforms the original UNet and its variants, with only 6% number of parameters.

Furthermore, improving medical image segmentation effectiveness largely involves the training dataset. As methodology matures across medical vision fields, enhancing algorithm robustness for the transition from experimental research to clinical application becomes increasingly crucial. Within the realm of data augmentation, the GAN was widely used by generating new samples. Luc et al. [13] firstly applied the generative adversarial network (GAN) to image segmentation, utilizing the generative network for segmentation models while training the adversarial network as a classifier. Singh et al. [14] introduced a conditional generation adversarial network (cGAN) for segmenting breast tumors within the designated region (ROI) in mammograms.

## 3. Method

UNet++ improves upon UNet with nested and dense skip connections but introduces overfitting risks due to its large parameter scale. To address this, we propose a UNet++-based whole-process framework for lung segmentation, as illustrated in Figure 1. The methodology includes three key components: (a) Lung CT Image Augmentation to preprocess images, enhancing quality and segmentation accuracy; (b) the UNet++ Segmentation Model, a robust framework tailored for biomedical image segmentation tasks; and (c) Model Parameter Tuning using experimental parameter callbacks to mitigate overfitting and optimize segmentation performance. This framework ensures precision and reliability in lung CT segmentation.

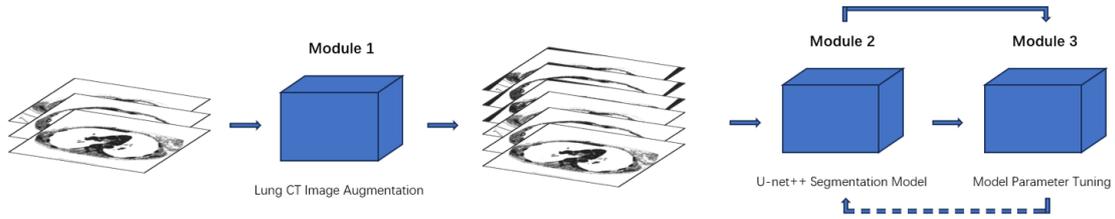

**Figure 1.** Pipeline of network

### 3.1. Lung Slice CT Image Augmentation

The CT images with signed masks dataset is small and simple. Oriented toward the problem, data augmentation implemented on CT images is of vital importance. The augmentation first applies specific images inputting method to maintain the original quality of images and masked images. The images fed into the program then are rotated randomly, enlarging the total amount of dataset twice. Alignment between images and masks ensures no error in the following sections. All of these newly generated data images are added into the whole dataset.

### 3.2. UNet++ Segmentation Model

Figure 3 shows the base batch of UNet++. Compared with UNet, UNet++ combines simple UNets of varying depths, and utilizes dense skip connections. Inherent from UNet's encoder-decoder architecture, these connections build relations between each sub-network; and as designed, the lower-level node functions as deep supervision within the sub-network. The network architecture of UNet++ is laid out, with skip pathways and dense connections shown in red and blue. Deep supervision is shown in yellow. Before outputting, an additional mask module designed for lung segmentation task is attached.

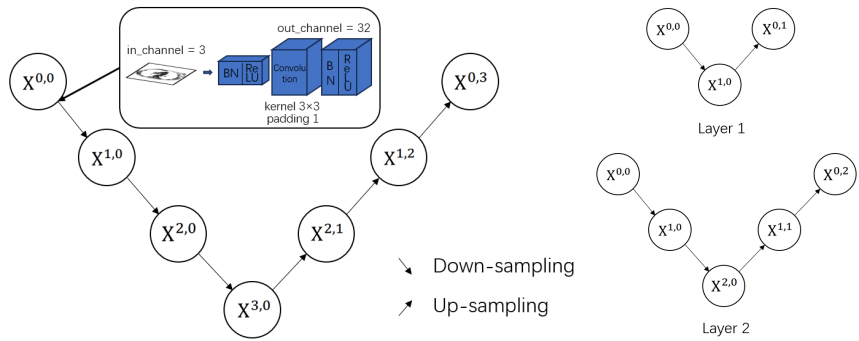

**Figure 3.** The base batch of UNet++

The main idea behind UNet++ is to bridge the semantic gap between the feature maps of the encoder and decoder prior to fusion. In other words, the convolution output of $X^{i,h}$ and the up-sampled output of $X^{i+1,j}$ ($i$: the down-sampling layer along the encoder; $j$: the maximum of convolution layer of the dense block along the skip pathway; $k$: the depth maximum of network; $i < k$, $h < j$) are fused directly through a concatenation layer (skip connections) that precedes each convolution layer. Essentially, the dense convolution block brings the semantic level of the encoder feature maps closer to that of the feature maps.

To formulate it, we make the following pre-assumptions: function $H(\cdot)$ is a convolution operation with a proposed activation function following; $D(\cdot)$ and $U(\cdot)$ denote a down-sampling layer and an up-sampling layer respectively; [ ] represents the concatenation layer (skip connections). When $j = 0$, the node only receives the down-sampling output from the previous encoder; when $j > 0$, the node receives the convolution output(s) along the convolution layer with the help of skip connections, as well as the up-sampled output from the lower decoder network. Consequently, the feature map at level $i, j$ is computed as:

$$X^{i,j} = \begin{cases} H\left(D(X^{i-1,j})\right), & j = 0 \\ H\left([X^{i,k}]_{k=0}^{j-1}, U(X^{i+1,j-1})\right), & j > 0 \end{cases} \quad (1)$$

Differing from the original version of UNet++, we apply model pruning to the model and add a full connection layer at the end. The parameter quantity of a single continuous convolution module is calculated as follows:

$$N_{conv} = C_{in} \times K^2 \times C_{out} \quad (2)$$
$$N_{bn} = 2 \times C_{out} \quad (3)$$

where $N_{conv}$ represents parameters for each convolution layer, $N_{bn}$ represents parameters for each batch normalization layer, $K$ is denoted as the kernel size, $C_{in}$ and $C_{out}$ are denoted as the number of input and output channels, respectively.

By applying model pruning, the resultant parameters for the entire model are 31% lower than previous model, demonstrating a remarkable progress in alleviating overfitting.

*3.3. Model Parameter Tuning*

We employed early stopping and parameter callbacks to optimize model training. Early stopping terminated training when accuracy improvements fell below a 0.05 threshold per epoch, preventing overfitting. Parameter callbacks dynamically evaluated the model on validation data, saving the best-performing parameters throughout training. These methods ensured the model achieved optimal performance by avoiding overtraining and preserving the most effective parameter configurations, enhancing generalization across diverse scenarios.

## 4. Results

*4.1. Datasets and Implementation*

The data in this dataset is representative of the clinical variability and challenges encountered in real-world clinical settings. As mentioned earlier, extensive data augmentation techniques are applied, where each image is subjected to random transformations, effectively doubling the dataset to 267 sets.

Mixing the original images and augmented images, we split the data into 10 parts evenly, nine of which are used for training, the rest one for evaluation. By utilizing K-fold cross-validation on the small dataset, the training result is significantly higher than the common dataset splitting method. As demonstrated in table 2, the accuracy after applying K-fold cross-validation is 96.82%, overpassing that before by 37.74% on our examination network. Several K values are assessed for better result.

**Table 1.** Test accuracies comparison between different data partitioning strategies and dataset preparation method.

| K | Accuracy | K | Accuracy |
|---|---|---|---|
| (not used) | 60.28% | 8 | 90.02% |
| 5 | 90.01% | 9 | 90.27% |
| 6 | 91.26% | 10 | 94.82% |
| 7 | 92.93% | 11 | 87.64% |

*4.2. Experimental results*

Based on the pre-test experiment, we trained our network on total 534 sets of CT scans using K-fold cross-validation as our dataset partitioning and training method. We used binary cross-entropy as loss function, employed Adam optimizer with starting learning rate at 0.001. Besides, scheduler reducing learning rate on plateau is adopted to maximally alleviate overfitting issue. When training after three epochs without improvement, learning rate would decrease, finally to a minimum of 0.00001.

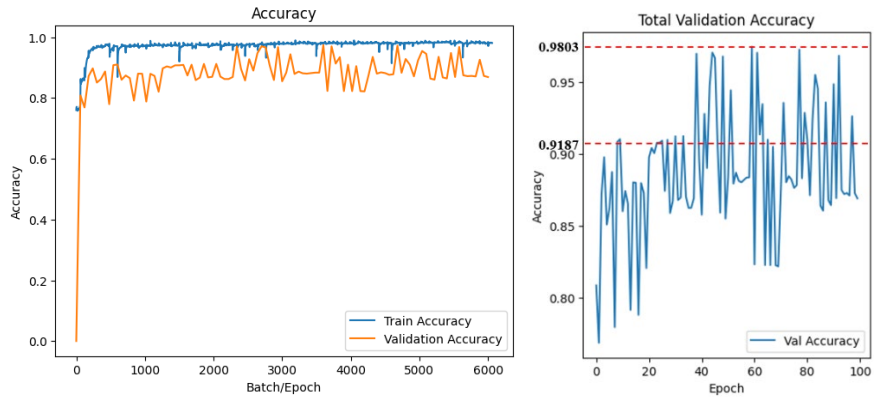

**Figure 4.** Training accuracy and validation accuracy

Index changing curves of train accuracy and validation accuracy are shown in figure 4. As indicated, our model convergent rapidly, especially on small range of dataset. Despite the fluctuations, we still reach an average accuracy of 91.87% after 400 epochs, outperforming a majority of other models in the task of segmenting lungs with the additionally lowest risk of overfitting. Distinct from other works, our network achieves the peak accuracy 98.03%. The training result were evaluated in terms of Dice overlap, computed as $0.9547 \pm 0.0145$. For intuitively showcasing our advancement, lung slice CT images processed by the network are displayed in couples as follows.

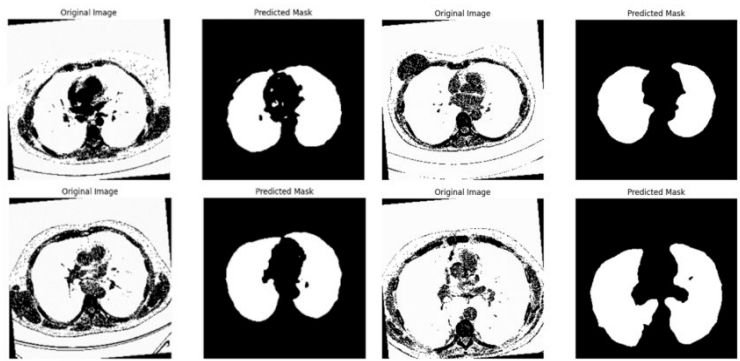

**Figure 5.** Segmentation results on test dataset

## 5. Conclusion

In this paper, we proposed a whole-process framework for segmentation of lung CT scans. The advanced performance is attributed to our comprehensive methods for mitigating overfitting. A novel deep learning network based on Unet++ is optimized in terms of network architecture and fine-tuned through parameter callbacks. In addition to the network architecture, image augmentation is introduced in order to minimize the influence of small dataset. By customizing our framework targeted on lung slice CT images, the results demonstrated highly practicability and effectiveness.

## 6. References

[1] Atun R, et al. (2015). "Expanding global access to radiotherapy." *The Lancet Oncology*, **16(10)**, pp. 1153–86. https://doi.org/10.1016/S1470-2045(15)00222-3
[2] Long J, Shelhamer E and Darrell T (2015). "Fully convolutional networks for semantic segmentation." *Proc. IEEE Conf. Comput. Vis. Pattern Recognit.*, pp. 3431–40. https://doi.org/10.1109/CVPR.2015.7298965
[3] Ronneberger O, Fischer P, Brox T (2015). "UNet: convolutional networks for biomedical image segmentation." *MICCAI 2015. LNCS*, vol. 9351, Navab N, et al. (Cham: Springer), pp. 234–41. https://doi.org/10.1007/978-3-319-24574-4_28
[4] M Drozdzal, E Vorontsov, G Chartrand, S Kadoury and C Pal (2016). "The importance of skip connections in biomedical image segmentation." *Deep Learning and Data Labeling for Medical Applications*, vol. 10008, Carneiro G, et al. (Cham: Springer), pp. 179–87. https://doi.org/10.1007/978-3-319-46976-8_19
[5] B Hariharan, P Arbelaez, R Girshick and J Malik (2015). "Hypercolumns for object segmentation and fine-grained localization." *Proc. IEEE Conf. Comput. Vis. Pattern Recognit.*, pp. 447–56. https://doi.org/10.1109/CVPR.2015.7298690
[6] Yuan Y, et al (2024). Deep learning for medical image segmentation, in *Machine Learning and Artificial Intelligence in Radiation Oncology*, J. Kang, et al. (Amsterdam: Academic Press), chapter 5, pp. 107-135.
[7] Oktay O, et al. (2018). "Attention UNet: learning where to look for the pancreas." Preprint at arXiv:1804.03999
[8] Ibtehaz N and Rahman M S (2020). "MultiResUNet: rethinking the UNet architecture for multimodal biomedical image segmentation." *Neural Netw.*, **121**, pp. 74–87. https://doi.org/10.1016/j.neunet.2020.05.012
[9] Z W Zhou, M M R Siddiquee, N Tajbakhsh and J M Liang (2018). "UNet++: A Nested UNet Architecture for Medical Image Segmentation." *Deep Learning in Medical Image Anylysis and Multimodal Learning for Clinical Decision Support*, pp: 3–11. https://doi.org/10.1007/978-3-030-00889-5_1
[10] M Z Khan, M K Gajendran, Y Lee and M A Khan (2021). "Deep Neural Architectures for Medical Image Semantic Segmentation: Review." IEEE Access, **9**, pp. 83002–24. https://doi.org/10.1109/ACCESS.2021.3086530
[11] Isensee F, Kohl Simon A A, Petersen Jens, Maier-Hein Klaus H (2021). "nnU-Net: a self-configuring method for deep learning-based biomedical image segmentation." *Nature Methods*. **18(2)**, pp. 203–11. https://doi.org/10.1038/s41592-020-01008-z
[12] Weng Y, Zhou T, Li Y, Qiu X (2019). "Nas-unet: neural architecture search for medical image segmentation." *IEEE Access*, **7**, pp. 44247–57. https://doi.org/10.1109/access.2019.2902457
[13] Luc P, Couprie C, Chintala S, Verbeek J (2016). "Semantic segmentation using adversarial networks." Preprint at arXiv:161108408
[14] Singh V K, et al. (2020). "Breast tumor segmentation and shape classification in mammograms using generative adversarial and convolutional neural net-work." *Expert Syst. Appl*, **139**, p. 112855. https://doi.org/10.1016/j.eswa.2019.112855